# Plasmascopy of ultrafast hot charge carriers

Dmitry A. Zimin,[1] Raja Sen,[2] Muhammad Qasim,[3] Jelena Sjakste,[4] Vladislav S. Yakovlev[3,5]

[1]*Laboratoty of Physical Chemistry, ETH Zürich, Vladimir-Prelog-Weg 2, 8049, Zürich, Switzerland.*
[2]*SATIE, CNRS, ENS Paris-Saclay, Université Paris-Saclay, Gif-sur-Yvette, 91190, France.*
[3]*Max-Planck-Institut für Quantenoptik, Hans-Kopfermann-Straße 1, 85748 Garching, Germany.*
[4]*Laboratoire des Solides Irradies, CEA/DRF/IRAMIS, École Polytechnique, CNRS, Institut Polytechnique de Paris, 91120 Palaiseau, France.*
[5]*Fakultät für Physik, Ludwig-Maximilians-Universität, Am Coulombwall 1, 85748 Garching, Germany.*

We demonstrate an electric field-resolved approach for probing ultrafast dynamics of photoinjected charge carriers in solids. Direct access to the electric field of few-cycle pulses enables us to measure a broadband response of a medium with associated plasma frequency. We prepare an ensemble of hot charge carriers with energies sufficient to trigger impact ionization and establish a framework to measure its dynamics. Our time-resolved study reveals the short-lived ultrafast dynamics consistent with impact ionization and carrier trapping. This approach provides a promising route for studying ultrafast many-body physics in photoexcited solids, with predictions from advanced theoretical models.

*Introduction*—Detailed understanding of hot charge carrier dynamics is crucial for advancing the performance of optoelectronic devices and enabling new functionalities, where the ability to manipulate and control hot charges becomes increasingly important [1]. These dynamics involve a complex interplay of interactions among electrons, holes, and phonons [2,3]. Impact ionization [4] is one of important consequences of this interplay with a central role in charge transport, energy dissipation, and device operation. For instance, carrier multiplication increases the light-conversion efficiency beyond the Shockley–Queisser limit [5]; in p–n junctions, it triggers the avalanche breakdown; in photodetectors, it is harnessed to achieve internal gain required for fiber-optic communications and low-light detection. These advantages however can be negated by opposing mechanisms that reduce the number of mobile charges.

Ultrafast photoinjection [6] invokes a complex sequence of processes [7] with distinct timescales. For highly photoexcited charges (typically, with excess energies exceeding the second conduction band minimum), the fastest processes are the loss of interband coherence due to electron-electron interactions and momentum scattering due to electron-phonon scattering. This typically occurs on a $\lesssim$10 femtoseconds (fs) [3,8,9] timescale, which is too fast for substantial energy transfer to the lattice. By the end of this stage, electrons and holes get redistributed in the reciprocal space [10], while their energy distributions are largely preserved, which is known as momentum equilibration [11,12]. Photoexcited electrons and holes may undergo impact ionization if the excess energy is higher than the material band gap. The carriers may also get trapped and undergo trap-assisted or Auger recombination [13]. The timescales of these competing processes depend on the charge and defect densities, as well as excess energies of carriers, leading to a transient nature of the overall effect. Therefore, an increase of the charge density due to the impact ionization may be present for a limited time until it is negated by trapping.

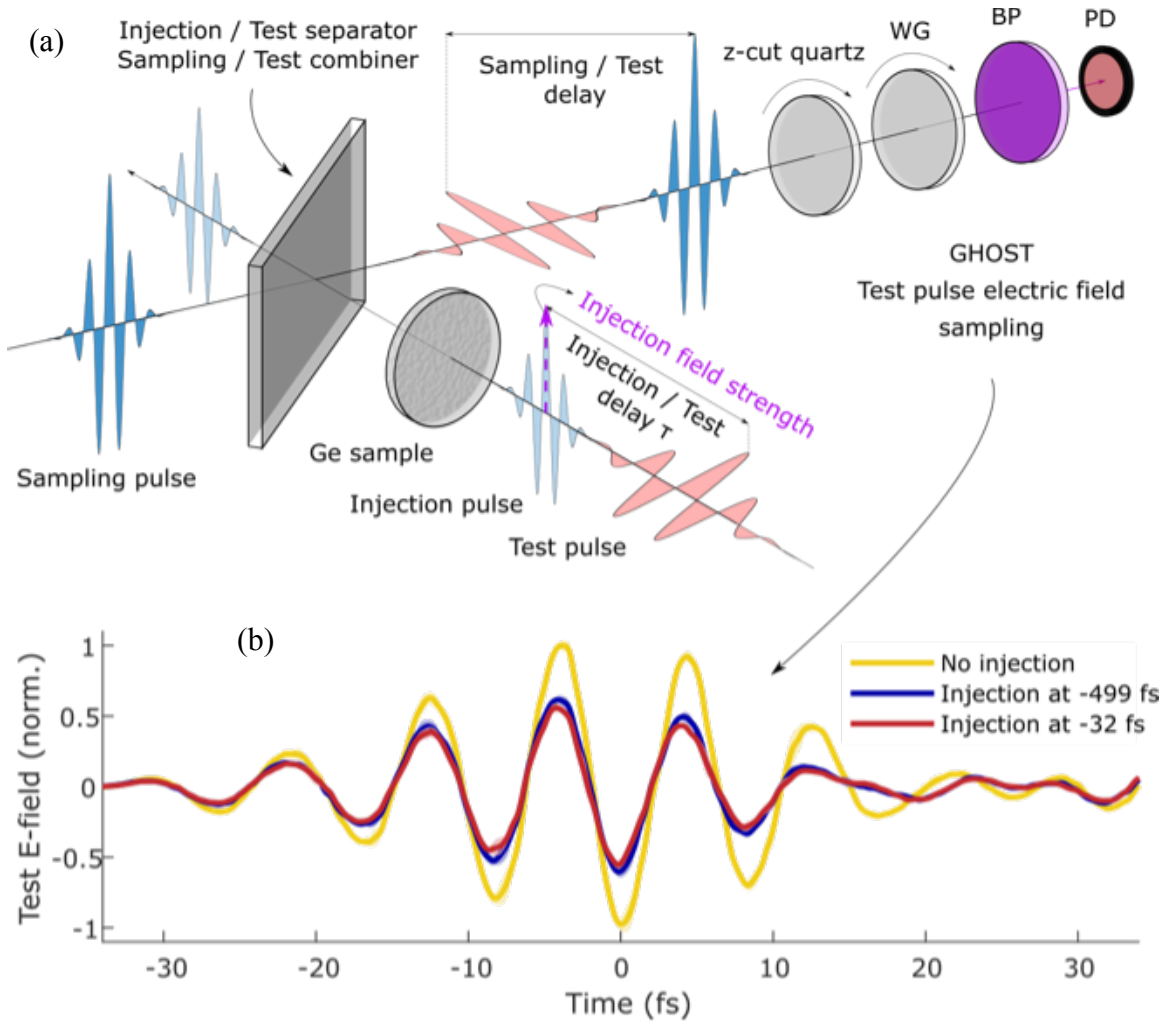

FIG 1. Experimental concept and results. (a) Schematic of the experiment. Two orthogonally polarized pulses (test pulse and injection pulse) are incident on a polycrystalline germanium. After interaction with the sample, the injection pulse is optically filtered out and the transmitted electric field of the test pulse is measured. The field-resolved measurement is repeated for various peak intensities of the injection pulse and various injection-test delays, $\tau$. (b) The measured waveform of the test pulse transmitted through the unperturbed sample (labeled "No injection") is compared to measurements where the sample is photoexcited by a 0.12 V/Å injection pulse at two different injection-test delays: the maximum delay in our scan ($\tau = 499$ fs) and the delay where photoexcitation occurs at the beginning of the test pulse ($\tau = 32$ fs). Here WG is a wire-grid polarizer, BP is a bandpass filter while PD is a photodiode. A typical measurement consisted of 70 injection-test delays and 3 measurements per delay [14].

These ultrafast dynamics are difficult to access experimentally. The time- and angle-resolved photoemission spectroscopy (tr-ARPES) provides the most direct information about occupations of Bloch states [3], however: the measurements require ultra-high vacuum; the short mean free path of the photoemitted electrons makes the technique surface-limited; the trade-off between the temporal and momentum resolutions, imposed by the uncertainty principle, puts a limit on the fastest accessible processes. The transient absorption [15,16] or reflection spectroscopy [17–19] provide only limited access since the valuable information imprinted on the temporal profile of light is lost.

Fieldoscopy [20] offers a direct access to the electric field preserving the full information. These measurements are standard to the terahertz (THz) domain, however the picosecond-scale duration of THz pulses is not sufficient to capture ultrafast femtosecond dynamics, while a narrow bandwidth limits the accuracy of the extraction of the complex frequency dependent material polarization. Recent advances in fieldoscopy have extended the bandwidth up to petahertz frequencies [21–28] and single photon sensitivity [29], enabling electric field-resolved studies with extreme sensitivity and sub-fs temporal resolution [6] across a broad bandwidth.

Here we present an electric field-resolved approach for accessing ultrafast scattering dynamics of photoexcited hot charge carriers. In contrast to THz studies, the response is imprinted on broadband few-cycle optical pulses. This allows us obtaining a broadband dynamic complex permittivity and associated plasma frequency ($\omega_{pl}$) with unprecedented temporal resolution. Since the plasma frequency

$$\omega_{\mathrm{pl}} = \sqrt{\frac{e^2 N}{\varepsilon_0 m_{\mathrm{eff}}}} \quad (1)$$

is governed by the charge density ($N$) and the effective mass ($m_{eff}$) [16,30], we use it as an observable to probe ultrafast charge dynamics in a photoexcited solid. Here $e$ is an elementary charge, $\varepsilon_0$ is the vacuum permittivity.

We focus on germanium, due to its low band gap that is favorable for many-body dynamics. We observe a transient evolution of the plasma frequency that indicates two competing mechanisms which we relate to the time-resolved observation of the ultrafast impact ionization in germanium opposed by charge trapping.

*Methods*—We realize a broadband field-resolved pump-probe study of hot charges with energies sufficient for impact ionization. We employ a recently developed GHOST [26] technique to measure, the electric field of a test pulse after transmission through a 290-nm [14] film of germanium (Fig. 1a). The small thickness is essential to avoid complex propagation effects. Like most free-standing samples of such a small thickness, our samples were polycrystalline.

The carrier-envelope phase (CEP) stable injection pulse with a duration of 2.7 fs [22] was used to photoexcite the medium and to confine the photoexcitation event to a temporal window shorter than hot-carrier relaxation time. The octave-spanning spectrum (1.24 – 2.48 eV) is well above the indirect (0.67 eV) and direct (0.8 eV) band gaps of germanium. Some carriers injected by this pulse have sufficient energy to create new electron-hole pairs via impact ionization (the lowest required photon energy is $2\times0.67=1.34$ eV). We varied the electric-field strength of the injection pulse between 0.05 V/Å and 0.15 V/Å (with no sample). Since the photoinjection is strongly nonlinear, the field inside the sample is smaller and cannot be directly related to the incident field by the linear Fresnel equations; the field values in our *ab initio* calculations (Fig. 3) therefore refer to the field inside the medium.

To record the non-equilibrium changes in the medium permittivity, we use a weak, CEP-stable, 12-fs test pulse of 2.1 μm central wavelength. Prior to photoinjection, germanium is transparent to the test spectrum (0.4 - 0.65 eV). The field strength was kept low (< 0.005 V/Å) to ensure the linearity of the polarization response to the test pulse. We record the transmitted test electric field for various injection-test delays and various field strengths of the injection pulse. Fig. 1b illustrates how photoinjection reshapes the test electric field.

*Results*—From the measured photoexcitation-induced changes in the test electric field, we calculate [14] the change in the complex permittivity (Fig. 2). Without photoinjection, $\varepsilon \approx 17$ for the test pulse. The emergence of electron-hole plasma decreases the real part of the permittivity, while the positive values of $\mathrm{Im}[\Delta\varepsilon]$ mean that the photoexcited sample absorbs light at frequencies where the band gap prohibits single-photon transitions in the unperturbed sample. Once electron-hole pairs recombine, the medium must return to its original state: $\Delta\varepsilon(\omega) \equiv 0$. From Fig. 2, we see that in our measurements, this process is still far from completion. The local maxima of $\mathrm{Im}[\Delta\varepsilon]$ at 0.47 eV and 0.56 eV indicate resonant interband transitions. According to our *ab initio* calculations (see Appendix), these are likely to be transitions between valence-band Bloch states around Γ point. These resonances are clearly visible even at $\tau$=499 fs with no notable change in their positions.

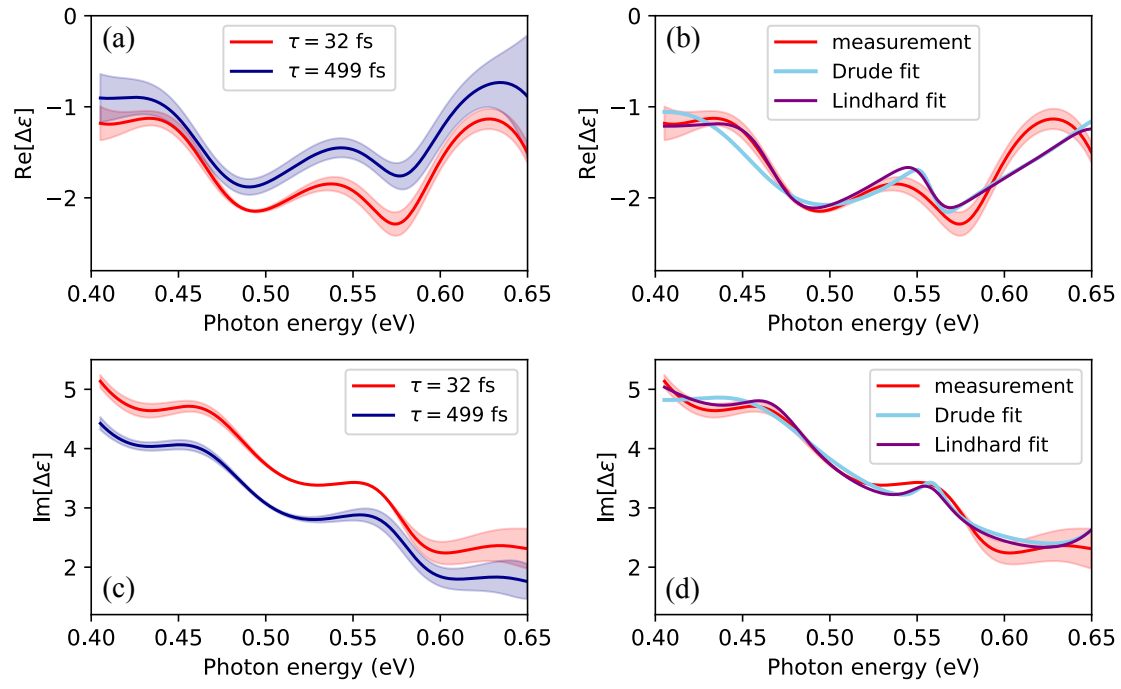

FIG 2. Photoinjection-induced changes in the linear permittivity of germanium. The 0.12 V/Å injection pulse changes the linear real (a, b) and imaginary (c, d) parts of Δε. The solid lines in (a) and (c) represent the permittivity change evaluated from the waveforms in Fig. 1b, while the shaded areas indicate one standard deviation. Panels (b) and (d) compare the measurements at $\tau = 32$ fs with the Drude-Lorentz and Lindhard-Lorentz models.

$$\Delta\varepsilon(\omega) = -\frac{\omega_{\mathrm{pl}}^2}{\left(\omega + i\gamma_{\mathrm{pl}}\right)^2} - \sum_l \frac{a_l}{\omega_l^2 - 2i\omega\gamma_l - \omega^2} \quad (2)$$

A direct comparison of Re[Δε] and Im[Δε] across the full fluence series is shown in Fig. 5 (see Appendix). We now focus on the created electron-hole plasma and associated plasma frequency, $\omega_{\mathrm{pl}}$. We retrieve $\omega_{\mathrm{pl}}$ by fitting the measured $\Delta\varepsilon(\omega)$ to a model that accounts for both the electron-hole plasma and resonant interband transitions. We find that the Lindhard-Lorentz model provides better agreement with our experiment than Drude-Lorentz approximation (Figs. 2b and 2c). Here, $\gamma_{\mathrm{pl}}$ represents the relaxation rate and partially accounts for the free-carrier absorption [31,32]. The sum on the right-hand side consists of Lorentz terms, each describing resonant absorption at a frequency $\omega_l$ with a damping rate $\gamma_l$, where $a_l$ controls the strength of the resonance. The key distinction between the Lindhard-Lorentz and Drude-Lorentz models lies in the first term: in the Drude model, it is $-\omega_{\mathrm{pl}}^2/\left(\omega^2 + i\gamma_{\mathrm{pl}}\omega\right)$. The Lindhard approximation has a more rigorous quantum-mechanical foundation as it accounts for the collective behavior of electrons in the random-phase approximation [33].

The same averaging over crystal momenta corresponds to the first stage of relaxation after photoinjection: long before the electrons and holes thermalize, electron-electron, electron-impurity, and short-wavelength (intervalley) electron-phonon scattering redistribute them in reciprocal space such that the occupation of any Bloch state depends only on the state's energy and not on its crystal momentum [34,35]. For our conditions, we estimate that if electron-phonon interaction were the only contributing process, momentum equilibration would take 11–33 fs for electrons and 8–9 fs for holes (see Appendix). The energy exchange between electrons and phonons is negligible during such a short time interval (1–2 eV/picosecond, see Appendix). Consequently, the energy distribution is expected to change negligibly.

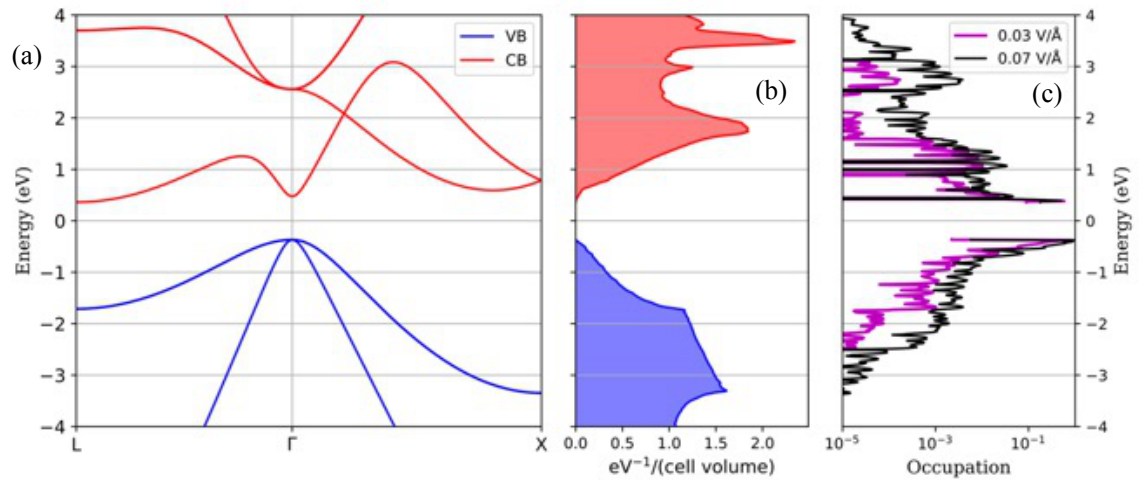

FIG 3. Photoexcitation of germanium by the injection pulse. (a) Band structure, (b) density of states, and (c) energy-resolved carrier occupations obtained from density functional theory. Each data point in panel (c) represents the occupation probability averaged over all Bloch states at a given energy. The calculations were performed for peak injection fields of 0.03 V/Å and 0.07 V/Å within the medium.

Fig. 3c shows that the injection pulse occupies states where the carrier energy is sufficient for impact ionization. Although these occupations are small, the high density of states at these energies makes impact ionization significant. The substantial reshaping of carrier occupations upon the increase in the field amplitude within the medium, from 0.03 V/Å to 0.07 V/Å, as well as the strong depletion of the states near the valence-band maximum, demonstrate the nonlinear character of photoinjection.

With these considerations we analyze the dependence of the measured $\omega_{\mathrm{pl}}$ (Eq. 1) on the field strength of the injection pulse $E_0$ and the time of photoinjection (Fig. 4a). Although $m_{\mathrm{eff}}$ increases with increasing $E_0$ [29,32], $\omega_{\mathrm{pl}}$ takes larger values for stronger injection because the increase in $m_{\mathrm{eff}}$ is offset by the increase in the electron-hole pairs concentration, $N$. What is more intriguing is the dependence of $\omega_{\mathrm{pl}}$ on time. The steady decrease of the $\omega_{\mathrm{pl}}$ for $t \gtrsim 100$ fs may look like an relaxation, but the electron-hole recombination is supposed to be negligible within our time interval [13,36]. Furthermore, as carriers lose their energies (e.g., by generating phonons) and gather at the bottoms of the L, Γ, and X valleys, where band curvature is largest, their optical effective mass must decrease [16], which would increase the plasma frequency. In the experiment, however, we observe an increase only during the first few tens of fs, after which $\omega_{\mathrm{pl}}$ decreases.

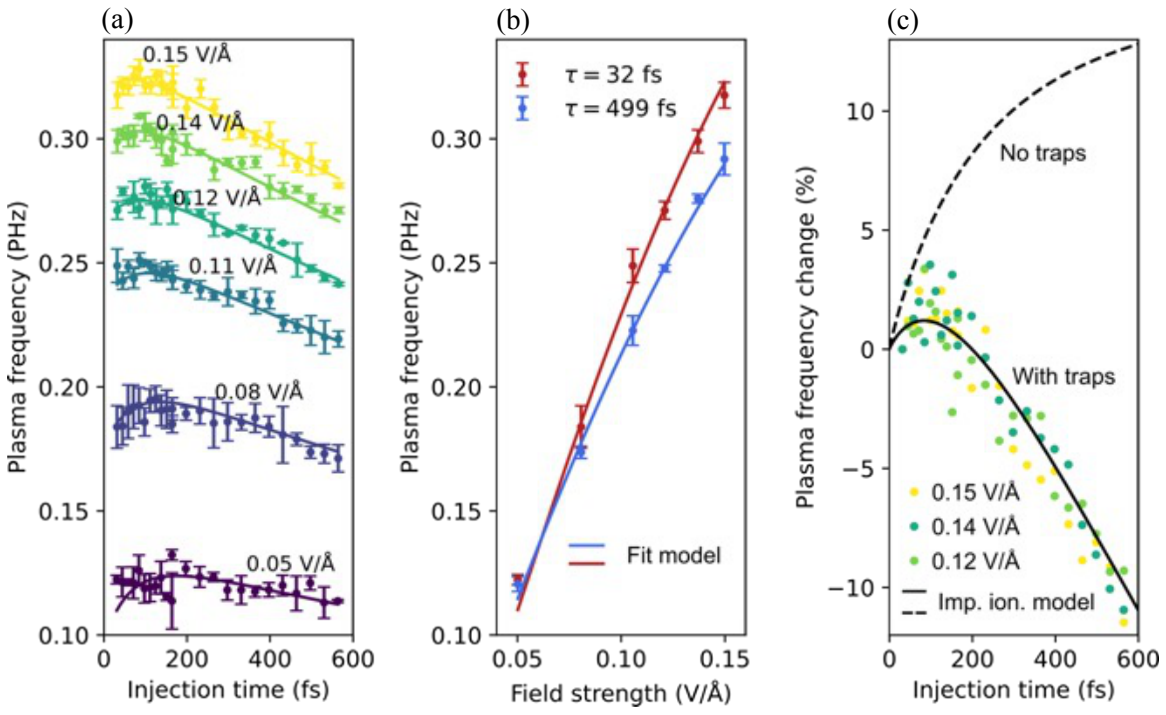

FIG 4. Plasma frequency dynamics. (a) Time evolution of the plasma frequency for different injection field strengths. Solid lines show fits using the model of Eq. (3). (b) Plasma frequency versus injection field strength at early ($\tau = 32$ fs) and late ($\tau = 499$ fs) times of photoinjection. Solid lines are obtained from the fit model of Eq. (3) utilized for panel (a). (c) Comparison of the experiment (dots) and theoretical estimation based on the impact ionization model. The dashed black curve represents the plasma frequency dynamic obtained from the ab initio calculated electron and hole populations and incorporating Monte Carlo–derived impact ionization rates for germanium [37]. The solid black curve shows the effect of charge trapping applied to the dashed black curve.

Auger recombination can be ruled out as a dominant cause of this decrease: its rate is proportional to $N^3$, so that when $E_0$ changes from 0.05 V/Å to 0.15 V/Å, the Auger rate must increase by ~ 27 times, whereas Fig. 4a shows no significant speed-up. We attribute the slow decrease primarily to carrier trapping, which is abundant in polycrystalline germanium [13]. Although the acceptor-defect concentration of pristine polycrystalline Ge films is between $10^{17}$ and $10^{19}$ $cm^{-3}$ [38], the traps are not restricted to the isolated acceptors: grain boundaries and dangling-bond defects provide additional trapping sites, so that the effective trap concentration can substantially exceed the net acceptor doping. For the strongest injection pulse (0.15 V/Å), we estimate $N \lesssim 2 \times 10^{20}$ $cm^{-3}$, so that even under the most extreme conditions a significant fraction of the photoinjected carriers can relax into trap states. An additional contribution comes from intervalley scattering from the bottom of the Γ valley, where the effective electron mass is $m^* = 0.05\, m_e$, to the bottom of the L valley, where the effective DOS mass is $m^* = 0.2\, m_e$ [34]: such a redistribution raises the effective mass and thereby lowers $\omega_{pl}$ at a fixed carrier density $N$. Although the single-particle Γ–L scattering time in intrinsic germanium is ~230 fs [39], it is extended to the picosecond scale at the carrier densities reached here [38]. We therefore attribute the slow decrease of $\omega_{pl}$ primarily to carrier trapping, with an additional contribution from a gradual increase of the effective mass (see Appendix).

A notable feature in Fig. 4a is the initial increase in $\omega_{pl}$ during the first few tens of fs, which becomes more prominent as the injection pulse gets stronger. We attribute it to the ultrafast impact ionization. This interpretation is consistent with theoretical predictions, as well as with experiments on narrow-gap semiconductors [41,42]. Photoinjected electrons with sufficient energy can excite valence electrons to the conduction band through inelastic collisions shortly after photoinjection, which increases $N$ and $\omega_{pl}$. Note that impact ionization also changes the effective mass. To confirm our considerations, we demonstrate that the measured $\omega_{pl}(t, E_0)$ can be approximated by a model that captures three key physical processes: impact ionization, carrier trapping, and the field-strength-dependent effective mass $m_{eff}$. We used the following model:

$$\omega_{pl}(t, E_0) = a\sqrt{\frac{E_0^2}{E_0^2 + E_1^2}\left(e^{-\gamma_1 t} - \frac{b\, e^{-\gamma_2 t}}{E_0^2 + E_2^2}\right)}. \quad (3)$$

The first term under the square root describes the slow decrease in $\omega_{pl}$ at large times, while the second term describes its increase shortly after photoinjection. The corresponding rates are $\gamma_1 = 5.6 \times 10^{-4}$ $fs^{-1}$ and $\gamma_2 = 0.02$ $fs^{-1}$, respectively, and they are independent of time and $E_0$. The model also captures the dependence of $\omega_{pl}$ on $E_0$ in the weak-field regime, as well as deviations from the $\omega_{pl}^2 \propto N \propto E_0^2$ scaling due to initial-state depletion, charge screening [43], and the dependence of the $m_{eff}$ on $E_0$. We account for these effects through intensity-dependent denominators in Eq. (3). The solid curves in Fig. 4A show the fit results with the following parameters: $a = 3.8$ $fs^{-1}$, $b = 0.013$, $E_1 = 0.23$ V/Å, and $E_2 = 0.049$ V/Å. While physically motivated, the model extracts the characteristic time scales: $\gamma_1^{-1} = 1.8$ ps for carrier trapping, which is consistent with [13], and $\gamma_2^{-1} = 50$ fs for the impact ionization.

While Fig. 4a emphasizes the dependence of $\omega_{pl}$ on time, Fig. 4b shows the dependence of $\omega_{pl}$ on the injection field strength for 32 fs and 499 fs times. This again indicates that the relaxation dynamics of hot charges differ significantly on timescales below and above 50 fs confirming two competing mechanisms. Overall, the fit model that considers two competing processes with $\gamma_2^{-1} = 50$ fs and $\gamma_1^{-1} = 1.8$ ps is consistent with our experiment.

*Discussion*—In our experiments on germanium, we observed that, after photoinjection by a few-fs laser pulse, the plasma frequency first increases during the first ~100 fs and then decreases on the picosecond time scale. We note that due to square root dependence of the plasma frequency, the 100 fs dynamics corresponds to 50 fs charge dynamics. We argue that the dominant contribution to the initial increase in $\omega_{pl}$ must come from impact ionization, which increases the carrier concentration and decreases the effective mass. The subsequent decrease in $\omega_{pl}$ at a rate too fast for electron-hole recombination implies a decrease in carrier concentration attributed to carrier trapping. To verify this, we perform *ab initio* calculation of the energy-dependent photoexcited charge density (see Appendix) and apply theoretical energy-dependent impact ionization rates derived from a Monte Carlo model [37]. We integrate over all energies to obtain the temporal evolution of the total charge density, incorporating charge trapping phenomenologically (see Appendix). Using Eq. (1), we estimate the time-dependent plasma frequency. Fig. 4c shows the result of our model together with the measured temporal changes in the plasma frequency at three highest injection field strength. Our estimation is based on the dynamic of the charge density only although the plasma frequency (Eq. 1) also depends on the effective mass. However, since the impact ionization redistributes charges in the momentum space, we expect impact ionization-induced change of the effective mass to be synchronized with the dynamic of the charge density. We find a good quantitative agreement of our impact ionization model with the experiment considering the trapping time constant of 1.27 ps. Fig. 4c highlights the importance of charge trapping in polycrystalline germanium explaining why the impact ionization effect is entirely suppressed by charge trapping after about 50 fs.

*Conclusions*—We have demonstrated a new approach for investigating ultrafast dynamics of hot charge carriers. The broadband nature of our experiment allows us to accurately extract the plasma frequency in spite of broad absorption peaks. Due to the square root dependence of $\omega_{pl}$ on the charge density and the effective mass we effectively trace fast dynamics of these quantities that are encoded in a slower observable — the plasma frequency — thereby relaxing experimental demands for accessing ultrafast processes. When the test pulse duration is sufficiently short compared to the timescale of changes in medium properties, the polarization response can be represented by the frequency-dependent, complex-valued permittivity—an approach we adopted in this work. However, a general formalism exists for describing the response of dispersive media with time-dependent properties [6].

We established a fit formula that together with our experiment suggest two competing mechanisms which we attribute to the impact ionization and charge trapping. To validate our findings, we conceived an impact ionization model and found a good agreement with the experiment, revealing, to our knowledge, the first time-resolved observation of the impact ionization in germanium. We find that this effect is short-lived as it is quickly overshadowed by charge trapping.

This approach is not limited to semiconductors and can be applied to other materials as long as the time-dependent electric field of the transmitted test pulse encodes the effect of the pump on the medium's optical properties. The concept provides a promising route for testing ultrafast, many-body physics in photoexcited solids with predictions from advanced theoretical models. For instance, Auger recombination or momentum equilibration [11,12] in conventional materials, or correlated effects in quantum materials.

## ACKNOWLEDGMENTS

This study is funded by the Swiss National Science Foundation (grants no. TMPFP2 217068 and PZ00-2_232911, D.A.Z.). J.S. and R.S. acknowledge ANR DragHunt project ANR-24-CE50-3505. Computer time has been granted by the national centers GENCI-CINES and GENCI-TGCC (Project 2210), and by École Polytechnique through the 3Lab computing cluster.

[1] Z. Chen et al., Ultrafast electron dynamics reveal the high potential of InSe for hot-carrier optoelectronics, Phys. Rev. B **97**, 241201 (2018).
[2] J. Shah, *Ultrafast Spectroscopy of Semiconductors and Semiconductor Nanostructures*, Vol. 115 (Springer Berlin Heidelberg, Berlin, Heidelberg, 1999).
[3] J. Sjakste, R. Sen, N. Vast, J. Saint-Martin, M. Ghanem, P. Dollfus, F. Murphy-Armando, and J. Kanasaki, Ultrafast dynamics of hot carriers: Theoretical approaches based on real-time propagation of carrier distributions, The Journal of Chemical Physics **162**, 061002 (2025).
[4] S. M. Sze, *Physics of Semiconductor Devices*, 2. ed., [Nachdr.] (Wiley, New York, N.Y., 2003).
[5] W. Shockley and H. J. Queisser, Detailed Balance Limit of Efficiency of p-n Junction Solar Cells, Journal of Applied Physics **32**, 510 (1961).
[6] D. A. Zimin, N. Karpowicz, M. Qasim, M. Weidman, F. Krausz, and V. S. Yakovlev, Dynamic optical response of solids following 1-fs-scale photoinjection, Nature **618**, 276 (2023).

[7] H. Haug and S. W. Koch, *Quantum Theory of the Optical and Electronic Properties of Semiconductors*, 5th ed. (WORLD SCIENTIFIC, 2009).
[8] J. Lloyd-Hughes et al., The 2021 ultrafast spectroscopic probes of condensed matter roadmap, J. Phys.: Condens. Matter **33**, 353001 (2021).
[9] M. Na, A. K. Mills, and D. J. Jones, Advancing time- and angle-resolved photoemission spectroscopy: The role of ultrafast laser development, Physics Reports **1036**, 1 (2023).
[10] H. Tanimura, J. Kanasaki, K. Tanimura, J. Sjakste, N. Vast, M. Calandra, and F. Mauri, Formation of hot-electron ensembles quasiequilibrated in momentum space by ultrafast momentum scattering of highly excited hot electrons photoinjected into the Γ valley of GaAs, Phys. Rev. B **93**, 161203 (2016).
[11] J. Sjakste, K. Tanimura, G. Barbarino, L. Perfetti, and N. Vast, Hot electron relaxation dynamics in semiconductors: assessing the strength of the electron–phonon coupling from the theoretical and experimental viewpoints, J. Phys.: Condens. Matter **30**, 353001 (2018).
[12] H. Tanimura, J. Kanasaki, K. Tanimura, J. Sjakste, and N. Vast, Ultrafast relaxation dynamics of highly excited hot electrons in silicon, Phys. Rev. B **100**, 035201 (2019).
[13] M. Zürch et al., Direct and simultaneous observation of ultrafast electron and hole dynamics in germanium, Nat Commun **8**, 15734 (2017).
[14] See Supplemental Material for details, which includes Refs. [52,53].
[15] M. Zürch et al., Ultrafast carrier thermalization and trapping in silicon-germanium alloy probed by extreme ultraviolet transient absorption spectroscopy, Structural Dynamics **4**, 044029 (2017).
[16] M. Wörle, A. W. Holleitner, R. Kienberger, and H. Iglev, Ultrafast hot-carrier relaxation in silicon monitored by phase-resolved transient absorption spectroscopy, Phys. Rev. B **104**, L041201 (2021).
[17] C. J. Kaplan, P. M. Kraus, A. D. Ross, M. Zürch, S. K. Cushing, M. F. Jager, H.-T. Chang, E. M. Gullikson, D. M. Neumark, and S. R. Leone, Femtosecond tracking of carrier relaxation in germanium with extreme ultraviolet transient reflectivity, Phys. Rev. B **97**, 205202 (2018).
[18] M. Neuhaus et al., Transient field-resolved reflectometry at 50–100 THz, Optica **9**, 42 (2022).
[19] G. Inzani et al., Field-driven attosecond charge dynamics in germanium, Nat. Photon. **17**, 1059 (2023).
[20] A. Srivastava, A. Herbst, M. M. Bidhendi, M. Kieker, F. Tani, and H. Fattahi, Near-petahertz fieldoscopy of liquid, Nat. Photon. **18**, 1320 (2024).
[21] S. Keiber, S. Sederberg, A. Schwarz, M. Trubetskov, V. Pervak, F. Krausz, and N. Karpowicz, Electro-optic sampling of near-infrared waveforms, Nature Photonics **10**, 159 (2016).
[22] S. Sederberg et al., Attosecond optoelectronic field measurement in solids, Nature Communications **11**, 430 (2020).
[23] D. Zimin, M. Weidman, J. Schötz, M. F. Kling, V. S. Yakovlev, F. Krausz, and N. Karpowicz, Petahertz-scale nonlinear photoconductive sampling in air, Optica **8**, 586 (2021).
[24] M. R. Bionta, F. Ritzkowsky, M. Turchetti, Y. Yang, D. Cattozzo Mor, W. P. Putnam, F. X. Kärtner, K. K. Berggren, and P. D. Keathley, On-chip sampling of optical fields with attosecond resolution, Nat. Photonics **15**, 456 (2021).
[25] S. B. Park, K. Kim, W. Cho, S. I. Hwang, I. Ivanov, C. H. Nam, and K. T. Kim, Direct sampling of a light wave in air, Optica **5**, 402 (2018).
[26] D. A. Zimin, V. S. Yakovlev, and N. Karpowicz, Ultra-broadband all-optical sampling of optical waveforms, Science Advances **8**, eade1029 (2022).
[27] Y. Liu, J. E. Beetar, J. Nesper, S. Gholam-Mirzaei, and M. Chini, Single-shot measurement of few-cycle optical waveforms on a chip, Nat. Photon. **16**, 109 (2022).
[28] A. Husakou, N. Karpowicz, V. S. Yakovlev, M. Ivanov, and D. A. Zimin, *Towards Multi-Petahertz All-Optical Electric Field Sampling*.
[29] D. A. Zimin, A. Ashoka, F. Reiter, and A. Rao, Fieldoscopy at the quantum limit, Light Sci Appl **15**, 110 (2026).
[30] K. Sokolowski-Tinten and D. Von Der Linde, Generation of dense electron-hole plasmas in silicon, Phys. Rev. B **61**, 2643 (2000).
[31] H. Y. Fan, W. Spitzer, and R. J. Collins, Infrared Absorption in n -Type Germanium, Phys. Rev. **101**, 566 (1956).
[32] S. C. Baker-Finch, K. R. McIntosh, D. Yan, K. C. Fong, and T. C. Kho, Near-infrared free carrier

absorption in heavily doped silicon, Journal of Applied Physics **116**, 063106 (2014).
[33] A. V. Andrade-Neto, Dielectric function for free electron gas: comparison between Drude and Lindhard models, Rev. Bras. Ensino Fís. **39**, (2016).
[34] H. Tanimura, K. Tanimura, and J. Kanasaki, Ultrafast relaxation of photoinjected nonthermal electrons in the Γ valley of GaAs studied by time- and angle-resolved photoemission spectroscopy, Phys. Rev. B **104**, 245201 (2021).
[35] K. Tanimura, H. Tanimura, and J. Kanasaki, Ultrafast dynamics of photoinjected electrons at the nonthermal regime in the intra- Γ -valley relaxation in InP studied by time- and angle-resolved photoemission spectroscopy, Phys. Rev. B **106**, 125204 (2022).
[36] E. Gaubas and J. Vanhellemont, Dependence of carrier lifetime in germanium on resisitivity and carrier injection level, Applied Physics Letters **89**, 142106 (2006).
[37] B. Ghosh, X. Wang, X.-F. Fan, L. F. Register, and S. K. Banerjee, Monte Carlo Study of Germanium n- and pMOSFETs, IEEE Trans. Electron Devices **52**, 547 (2005).
[38] K. Nozawa, K. Igura, T. Mizoguchi, N. Saitoh, N. Yoshizawa, T. Suemasu, and K. Toko, Hydrogen passivation effects on polycrystalline germanium thin films, NPG Asia Mater **17**, 26 (2025).
[39] G. Mak and H. M. Van Driel, Femtosecond transmission spectroscopy at the direct band edge of germanium, Phys. Rev. B **49**, 16817 (1994).
[40] F. Murphy-Armando, É. D. Murray, I. Savić, M. Trigo, D. A. Reis, and S. Fahy, Electronic heat generation in semiconductors: Non-equilibrium excitation and evolution of zone-edge phonons via electron–phonon scattering in photo-excited germanium, Applied Physics Letters **122**, 012202 (2023).
[41] H. Tanimura, J. Kanasaki, and K. Tanimura, Ultrafast scattering processes of hot electrons in InSb studied by time- and angle-resolved photoemission spectroscopy, Phys. Rev. B **91**, 045201 (2015).
[42] S. Biasco, F. Burri, S. Houver, E. Abreu, M. Savoini, and S. L. Johnson, Impact ionization in low-band-gap semiconductors driven by ultrafast terahertz excitation: Beyond the ballistic regime, Phys. Rev. B **106**, 235201 (2022).
[43] T. Sjodin, H. Petek, and H.-L. Dai, Ultrafast Carrier Dynamics in Silicon: A Two-Color Transient Reflection Grating Study on a ( 111 ) Surface, Phys. Rev. Lett. **81**, 5664 (1998).
[44] R. Kienberger et al., Atomic transient recorder, Nature **427**, 817 (2004).
[45] F. Tran and P. Blaha, Accurate Band Gaps of Semiconductors and Insulators with a Semilocal Exchange-Correlation Potential, Phys. Rev. Lett. **102**, 226401 (2009).
[46] The Elk Code, (2025).
[47] M. Qasim, D. A. Zimin, and V. S. Yakovlev, Optical Gain in Solids after Ultrafast Strong-Field Excitation, Physical Review Letters **127**, (2021).
[48] M. Qasim, M. S. Wismer, M. Agarwal, and V. S. Yakovlev, Ensemble properties of charge carriers injected by an ultrashort laser pulse, Phys. Rev. B **98**, 214304 (2018).
[49] P. Giannozzi et al., Advanced capabilities for materials modelling with Quantum ESPRESSO, J. Phys.: Condens. Matter **29**, 465901 (2017).
[50] G. Pizzi et al., Wannier90 as a community code: new features and applications, J. Phys.: Condens. Matter **32**, 165902 (2020).
[51] S. Poncé, E. R. Margine, C. Verdi, and F. Giustino, EPW: Electron–phonon coupling, transport and superconducting properties using maximally localized Wannier functions, Computer Physics Communications **209**, 116 (2016).
[52] T. Amotchkina, M. Trubetskov, D. Hahner, and V. Pervak, Characterization of e-beam evaporated Ge, $YbF_3$ , ZnS, and $LaF_3$ thin films for laser-oriented coatings, Appl. Opt. **59**, A40 (2020).
[53] T. N. Nunley, N. S. Fernando, N. Samarasingha, J. M. Moya, C. M. Nelson, A. A. Medina, and S. Zollner, Optical constants of germanium and thermally grown germanium dioxide from 0.5 to 6.6eV via a multisample ellipsometry investigation, Journal of Vacuum Science & Technology B, Nanotechnology and Microelectronics: Materials, Processing, Measurement, and Phenomena **34**, 061205 (2016).

## END MATTER

## APPENDIX: EXPERIMENTAL DESIGN

The experiments were performed with a beamline that delivers CEP-stable 2.7-fs optical pulses centered at about 750 nm wavelength [6]. Pulses from a Ti:Sapphire oscillator (Femtolasers) were amplified in a chirped-pulse amplifier at a repetition rate of 3 kHz, delivering pulses of 21 fs duration with 2.5 W average power. The pulses were spectrally broadened in a hollow-core fiber and further compressed in a multi-pass chirped mirror compressor (Ultrafast Innovations) to about 2.7 fs [22] duration (full width at half-maximum of intensity). The pulses were characterized in previous experiments using attosecond streaking [44] and non-linear photoconductive sampling [22]. The pulses were split into three optical arms (more detail in [6]), named as sampling, test, and injection. The test arm was down-converted in spectrum via intrapulse difference-frequency generation (iDFG) in a type II BiB3O6 (BiBo) crystal. The generated pulses centered at about 130 THz were further compressed with a combination of thin silicon and fused silica plates providing 12 fs (full width at half-maximum of intensity) duration. Silicon plates were also used to block the fundamental spectrum after the iDFG process. The pulses were further recombined using broadband metallic wire-grid polarizers, while pulse energy in each arm was set by additional thin wire grid polarizers in each arm. The transmitted test pulse waveform was measured with a GHOST metrology [26] using ~12-µm-thick z-cut α-quartz crystal. From this measurement, we estimate the peak electric field of the test pulse to be below 0.005 V/Å. The measurement of reference and modified (due to the excitation arm) waveforms was conducted with semi-simultaneous detection scheme [6] for high sensitivity.

## APPENDIX: ULTRAFAST PHOTOINJECTION MODEL

We modeled the interaction of a 3.5-fs, 800-nm pump pulse with a germanium crystal by numerically solving the time-dependent Schrödinger equation (TDSE) in the basis of Kohn-Sham states obtained from density functional theory (DFT). We used the Tran-Blaha modified Becke-Johnson exchange-correlation pseudopotential [45] in the Elk code [46]. The DFT calculations provided energies and momentum-operator matrix elements for 14 bands below the Fermi energy and 41 conduction bands on a 32×32×16 Monkhorst-Pack k-grid, yielding a direct band gap of 0.84 eV. Using this data, we solved the TDSE in the velocity gauge as described in [47]. To generate Fig. 3C, we modeled momentum equilibration [12] neglecting inelastic scattering. Under the assumption that carriers maintain their energies during redistribution in reciprocal space, we calculated the equilibrated probabilities $p(\epsilon)$ for electron occupation of a Bloch state at energy $\epsilon$ using conservation of carrier number and energy. For this calculation, we used the following equation as a starting point:

$$\int_{-\infty}^{\epsilon} d\epsilon' p(\epsilon')\,\mathrm{DOS}(\epsilon') = \frac{2}{N_k} \sum_{\epsilon_n(\boldsymbol{k})\leq\epsilon} f_n(\boldsymbol{k})\,, (4)$$

where DOS is the density of states, $N_k$ is the total number of k-grid nodes, $f_n(k)$ is the occupation of the Bloch state in band n at crystal momentum $k$, and the factor 2 accounts for spin degeneracy due to the absence of spin-orbit interaction in our DFT calculations. This approach allowed us to perform averaging without introducing energy bins, which would be problematic because, even with $N_k = 16384$, only a few Bloch states will be found in a few-meV energy interval. Instead, we evaluated the right-hand side of Eq. (4) and then used the Savitzky-Golay filter with third-order polynomials to evaluate the derivative with respect to $\epsilon$, which resulted in $p(\epsilon)$ $DOS(\epsilon)$. With the known density of states, we then calculated $p(\epsilon)$ and plotted it in Fig. 3C.

We also performed ab initio calculations of Δε(ω) using the model from [48]. In these simulations, we observed absorption peaks similar to the ones in Fig. 2. These resonances formed at 0.48 eV, 0.60 eV, and 0.63 eV, showing a small deviation from the measured resonances at 0.47 eV and 0.56 eV. This discrepancy likely originates from the neglected spin-orbit interaction. For the 0.48-eV resonance, we identified its origin as the excitation of holes from the light-hole band to the next lower-energy band at $k = 0.052$ atomic units (the reciprocal lattice vectors have a length of 1.02 atomic units).

To calculate the electron-phonon scattering times, germanium was described within density functional theory and density functional perturbation theory (DFPT) using Quantum ESPRESSO [49], Wannier90 [50] and EPW [51] codes. Computational details for Ge can be found in [3].

## APPENDIX: IMPACT IONIZATION MODEL

After obtaining the energy dependent occupations of electrons [$p_e(\epsilon)$] and holes [$p_h(\epsilon)$] followed by the photoinjection, we evaluate energy dependent impact ionization rates for electrons [$\gamma_h(\epsilon)$] and holes [$\gamma_h(\epsilon)$] with a previously reported Monte Carlo model [37]. We then simulate the temporal dependence of the total charge density ($N$) as follows:

$$N_e(t) = \int_0^{\infty} p_e(\epsilon) + \left(1 - e^{-\gamma_e(\epsilon)t}\right) p_e(\epsilon) d\epsilon,$$

$$N_h(t) = \int_{-\infty}^{0} p_h(\epsilon) + \left(1 - e^{-\gamma_h(\epsilon)t}\right) p_h(\epsilon) d\epsilon,$$

$$N(t) = N_e(t) + N_h(t).$$

To phenomenologically account for the trapping of charge carriers we assume that the charge density ($N$) exponentially decreases with a characteristic time constant $\tau$:

$$N(t) = N(t)e^{-t/\tau}.$$

Finally, we estimate the temporal dependence of the plasma frequency on the charge density as

$$\omega_{\text{pl}}(t) \propto \sqrt{N(t)}$$

## APPENDIX: SLOW DECREASE OF THE PLASMA FREQUENCY

The slow, picosecond-scale reduction of $\omega_{\text{pl}}$ is governed primarily by carrier trapping, which is abundant in polycrystalline germanium [13]. The acceptor-defect concentration in polycrystalline Ge is $10^{17}$–$10^{19}$ $\text{cm}^{-3}$ [38]; because grain boundaries and dangling-bond defects provide further trapping sites in addition to the isolated acceptors, the effective trap density can substantially exceed acceptor doping, although we cannot reliably quantify it. A gradual increase of the effective mass provides an additional contribution: a redistribution of the photoexcited electrons from the light Γ valley ($m^* \approx 0.05, m_e$) to the heavier L valleys ($m^* \approx 0.2, m_e$) raises the optical effective mass and lowers $\omega_{\text{pl}}$ at a fixed carrier density $N$ [34]. Although the single-particle Γ–L intervalley scattering time in Ge is only ∼ 230 fs [39], first-principles calculations combined with time-resolved x-ray scattering show that, at carrier densities of order $10^{20}$ $\text{cm}^{-3}$, the coupled electron–phonon system remains out of equilibrium for several picoseconds: strongly driven zone-edge phonons with low group velocities accumulate and are reabsorbed by the carriers, slowing the net intervalley redistribution to the picosecond scale [40]. Because both mechanisms lower $\omega_{\text{pl}}$ on comparable timescales, we cannot rigorously separate their contributions; the phenomenological model of Eq. (3) represents the net decay through a single trapping rate $\gamma_1$.

Finally, we note that our observable is the photoinjection-induced change of the permittivity, $\Delta\varepsilon(\omega)$, which vanishes in the absence of the injection pulse. The free-carrier response of the equilibrium (dopant) carriers is therefore contained in the unperturbed permittivity and does not enter $\Delta\varepsilon$; the retrieved plasma frequency thus reflects the photoinjected plasma. At long delays, as photoinjected carriers are trapped, $\Delta\varepsilon$ and the retrieved $\omega_{\text{pl}}$ relax toward zero rather than toward a floor set by the equilibrium carriers.

## APPENDIX: FLUENCE DEPENDENCE OF THE DIELECTRIC RESPONSE

Fig. 5 compares the transient permittivity measured at different injection field strengths at a fixed pump–probe delay $\tau \approx 32$ fs, close to the maximum of $\omega_{\text{pl}}$. The left column shows the measured $\text{Re}[\Delta\varepsilon]$ and $\text{Im}[\Delta\varepsilon]$, and the right column shows the corresponding Lindhard-Lorentz fits. Across this series the extracted plasma frequency increases from $\omega_{\text{pl}}/2\pi = 0.12$ PHz to 0.32 PHz. Both $|\text{Re}[\Delta\varepsilon]|$ and $|\text{Im}[\Delta\varepsilon]|$ grow monotonically with the injection field and are separated by more than the measurement uncertainty, so the nearly threefold change in $\omega_{\text{pl}}$ is directly reflected in the dielectric response.

The Lindhard-Lorentz model reproduces the fluence dependence and the overall spectral shape, but the fits are not perfect: the fitted curves deviate visibly from the data, particularly near the edges of the probed spectral window. This indicates that a Lindhard-Lorentz (or Drude-Lorentz) description, while adequate for extracting $\omega_{\text{pl}}$, does not fully capture the complexity of the hot-carrier dielectric response; a microscopic description of $\Delta\varepsilon(\omega)$ remains a subject for future work.

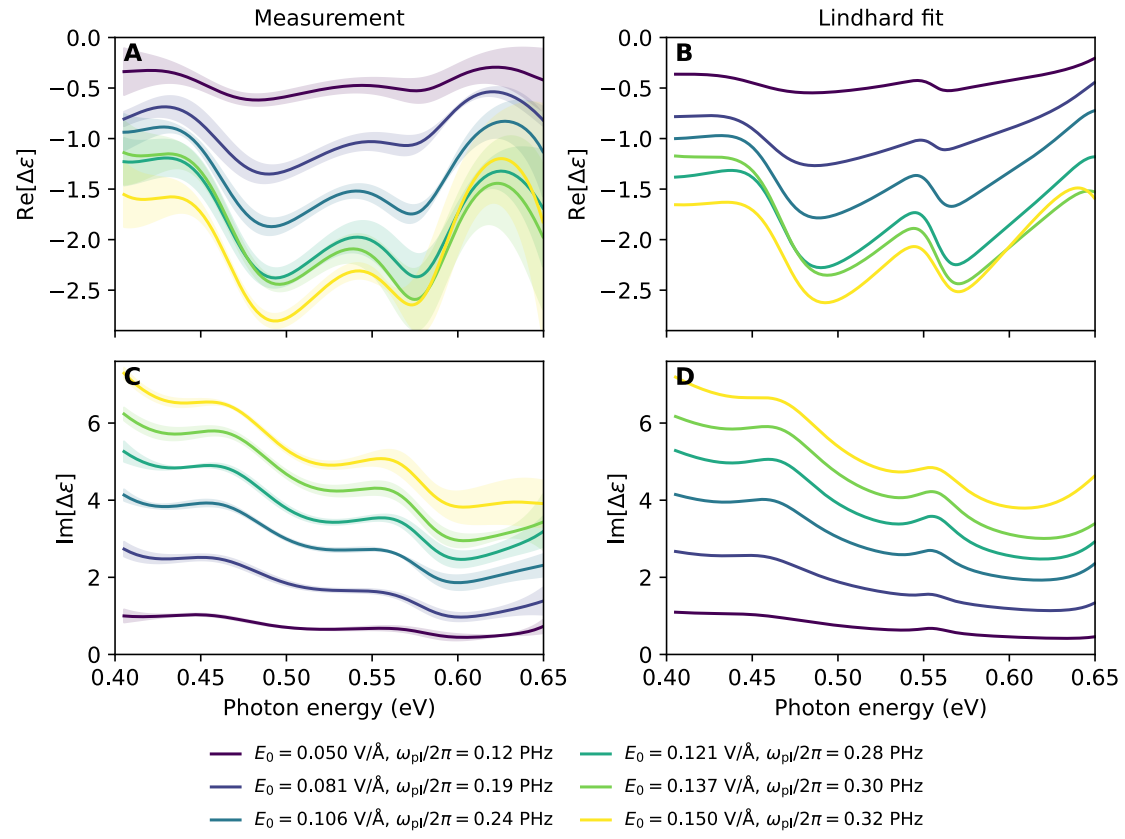

FIG 5. Fluence dependence of the transient dielectric response. Measured (left) and Lindhard-Lorentz–fitted (right) real (top) and imaginary (bottom) parts of $\Delta\varepsilon$ at $\tau \approx 32$ fs, for six injection field strengths $E_0$ (color scale; the corresponding plasma frequencies are given in the legend). Shaded bands in the measurement panels indicate $\pm 1\sigma$ over three scans. The weakest injection setting, dominated by noise, is omitted.